# Implementing Deep Reinforcement Learning-Based Grid Voltage Control in Real-World Power Systems: Challenges and Insights

Di Shi[1], Qiang Zhang[2], Mingguo Hong[2], Fengyu Wang[1], Slava Maslennikov[2], Xiaochuan Luo[2], Yize Chen[2]
[1]Klipsch School of Electrical & Computer Engineering, New Mexico State University, NM
[2]Department of Advanced Solution Technologies, ISO New England, Holyoke, MA
Email: dshi@nmsu.edu, xluo@iso-ne.com

*Abstract*—Deep reinforcement learning (DRL) holds significant promise for managing voltage control challenges in simulated power grid environments. However, its real-world application in power system operations remains underexplored. This study rigorously evaluates DRL's performance and limitations within actual operational contexts by utilizing detailed experiments across the IEEE 14-bus system, Illinois 200-bus system, and the ISO New England node-breaker model. Our analysis critically assesses DRL's effectiveness for grid control from a system operator's perspective, identifying specific performance bottlenecks. The findings provide actionable insights that highlight the necessity of advancing AI technologies to effectively address the growing complexities of modern power systems. This research underscores the vital role of DRL in enhancing grid management and reliability.

## I. Introduction

Deep reinforcement learning (DRL) is poised to catalyze a transformative shift in power systems control, uniquely equipped to handle the multifaceted decision-making challenges inherent in modern grid operations [1]-[5]. Compared to traditional control methods, DRL offers substantial advantages in navigating the intricate nonlinear dynamics of these complex systems. Its efficacy in simulated environments is well-documented, suggesting a broad potential that spans from real-time grid management to strategic foresight in energy market planning [6]-[10].

Despite these promising advances, the practical deployment of DRL in operational power systems remains nascent, with significant hurdles such as managing incomplete datasets, necessitating real-time adaptation to evolving system behaviors, and ensuring consistent reliability during network disturbances [11]. The literature reveals a pronounced gap in empirical research on DRL's deployment in complex real-world settings, underscoring the need for further investigation.

This study aims to bridge this gap by rigorously evaluating DRL's capability for operational control within realistic environment. Specifically, we address the following research questions: 1) how does DRL perform in real-world power system operations compared to its performance in simulated environments? 2) what are the impacts of model inaccuracies on the performance of DRL agents in operational contexts? 3) how adaptable are DRL agents to changes in network topology, such as N-1 and N-2 contingencies? 4) can DRL agents generalize to manage previously unencountered scenarios in real-world operations? 5) what are the practical implications of deploying DRL for voltage control in terms of system efficiency and resource conservation?

To answer these questions, we conduct methodical trials using established test models such as the IEEE 14-bus and Illinois 200-bus systems, and the production-grade ISO New England node-breaker model [12]-[15]. Our analysis critically assesses DRL's effectiveness for grid control from a system operator's perspective, identifying specific performance bottlenecks. The findings provide actional insights that highlight the necessity of advancing AI technologies to effectively address the growing complexities of modern power systems [16]-[19].

Addressing these research questions has significant implications for the future of power system operations. By understanding the performance of DRL in real-world settings, we can identify potential improvements and modifications required for practical deployment. Investigating the impacts of model inaccuracies and the adaptability of DRL agents to network changes ensures the robustness and reliability of these technologies. Furthermore, examining the generalization capabilities of DRL agents and their practical implications for voltage control can lead to more efficient, resilient, and sustainable power grid management. These insights not only advance academic research but also provide practical guidance for system operators and policymakers aiming to integrate AI-driven solutions into the energy sector.

By integrating technological innovation with the practicalities of operational practices, this study addresses a significant gap in existing research. We provide a novel synthesis of theoretical and applied knowledge, exploring DRL within the context of power system operations. The following sections will detail DRL's principles, algorithmic nuances, and practical applications through case studies to offer a comprehensive overview of its capabilities and limitations and in its integration into grid control systems.

## II. Formulations and Algorithms

### A. DRL Basics

DRL optimizes cumulative rewards through sequential interactions with an environment, as conceptualized in Fig. 1. At each timestep $t$, an agent observes a state $s_t$, selects an action $a_t$ based on policy $\pi$, and receives an immediate reward $r_t$. The environment then transitions to the next state $s_{t+1}$. This process generates state-action-reward sequences that inform the agent' policy development.

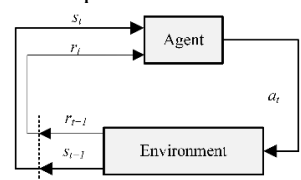

Fig. 1 The DRL framework.

*This work was supported in part by ISO New England, and the National Science Foundation (NSF) under Award Nos. 2301938 and 2418359.*

The DRL framework is structured around a state space $S$, an action space $A$, transition dynamics $T(s_{t+1}|s_t, a_t)$, a reward function $R(s, a, r)$, and a discount factor $\gamma \in [0, 1]$ which collectively define a Markov decision process (MDP) as $<S, A, T, R, \gamma>$.

In DRL-based Autonomous Voltage Control (AVC), episodes start with various grid conditions, drawing on data from SCADA or WAMS. The agent's goal is to correct voltage violations, with rewards reflecting the action's effectiveness. An episode concludes when control objectives are met, or a predefined number of iterations is reached. This method refines the agent's voltage regulation proficiency.

The reward function is critical in guiding agent behavior through control iteration. The reward for the $i^{th}$ control iteration, denoted as $r_i$, is typically calculated based on the voltage magnitude at bus $V_i$, with predefined thresholds distinguishing between normal operation, violation, and divergence zones. For instance, a voltage within the normal range (0.95-1.05 p.u.) might yield a positive reward $R_p$, while a voltage outside this range could incur a negative reward -$R_n$. In cases where the power flow diverges or the voltage profiles are too bad, a substantial penalty $R_{penalty}$ is applied. The following equation (1) gives an example of the reward definition, where $F_p$ and $F_n$ are additional terms used to shape the reward, which will be discussed in section III.B.

$$r_i = \begin{cases} R_p + F_p, & \forall V_j \in [0.95, 1.05] \\ -R_n + F_n, & \forall V_j \in [0.8, 1.2] \text{ and } \exists V_j \notin [0.95, 1.05] \\ R_{penalty}, & \text{otherwise} \end{cases} \quad (1)$$

The final return $r_f$ for an episode, comprising $n$ iterations, is expressed as the average of individual rewards:

$$r_f = \sum_1^n r_i / n; \quad (2)$$

This reward shaping mechanism refines agent actions toward maintaining desired voltage levels, enhancing the learning efficiency. It is important to note that there are numerous ways to define a reward function; the structure discussed here is just one example. Depending on the specific operational objectives, additional terms could be incorporated to optimize for system losses, control efforts, or other performance metrics. Such flexibility in reward function design allows customization to meet the unique requirements of different power system applications.

*B. DRL Algorithms Examined*

Our work assesses three DRL algorithms: Deep Q-Network (DQN), Deep Deterministic Policy Gradient (DDPG), and Soft Actor-Critic (SAC), each with unique operational strengths and constraints [1], [11], and [19].

DQN, which addresses problems with discrete action spaces, learns the optimal action-value function $Q^*(s,a)$ using a deep neural network as an approximator. It updates the Q-values following the Bellman equation, accounting for the learning rate $\alpha$ and the discount factor $\gamma$:

$$Q_{new}^{(s_t, a_t)} \leftarrow Q^{(s_t, a_t)} + \alpha [r_{t+1} + \gamma max_a Q^{(s_{t+1}, a)} - Q^{(s_t, a_t)}]. \quad (3)$$

DDPG caters to continuous action spaces and learns a Q-function and a policy simultaneously. It updates the policy network parameters $\theta^\mu$ using a policy gradient method and refines the Q-function similarly to DQN but informed by the policy's action outputs. It utilizes the following policy gradient update for the policy $\mu(s|\theta^\mu)$:

$$\nabla_{\theta^\mu} J \approx \frac{1}{N} \sum \nabla_a Q^{(s,a)}|_{s=s_i, a=\mu(s_i)} \nabla_{\theta^\mu} \mu(s|\theta^\mu)|_{s_i} \quad (4)$$

DDPG updates the Q-function as below.

$$Q_{new}^{(s_t, a_t)} \leftarrow Q^{(s_t, a_t)} + \alpha [r_{t+1} + \gamma max_a Q^{(s_{t+1}, \mu(s_{t+1}))} - Q^{(s_t, a_t)}]. \quad (5)$$

During the exploration process, the exploration policy $\mu'$ is designed by adding a random decaying noise $\xi$ as:

$$\mu'(s_i) = \mu(s_i|\theta^\mu) + \xi_i, \quad (6)$$

where $\xi_{i+1} = r_d \times \varepsilon_i$.

SAC, an off-policy algorithm, optimizes a stochastic policy in an entropy-regularized framework, introducing an entropy term to the reward function to promote exploration. It aims to maximize a composite objective function, $J(\pi)$,

$$J(\pi) = \sum_t E_{(s_t, a_t) \sim \rho_\pi} \left[ \gamma^t \left( r^{(s_t, a_t)} + \alpha \mathcal{H}(\pi(\cdot|s_t)) \right) \right], \quad (7)$$

where $\mathcal{H}$ is the entropy of the policy $\pi$ and $\alpha$ controls the importance of the entropy term. The SAC critic update is performed using a soft Bellman backup across two critic networks, fostering stability:

$$Q_{new}^{(s_t, a_t)} \leftarrow E_{(s_t, a_t) \sim \rho_\pi} [r^{(s_t, a_t)} + \gamma (min_{i=1,2} Q_{\theta_i'}^{(s_{t+1}, a_{t+1})} + \alpha \mathcal{H}(\pi(a_{t+1}|s_t)))], \quad (8)$$

where the minimum operation over $i=1, 2$ indicates that the soft update is taken over both critic networks to provide a more stable target.

### III. CASE STUDY AND DISCUSSION

*A. Benchmark System Test Cases*

To evaluate the performance of DRL-based AVC agents, our study employs three distinct testing systems of varying complexity:

1. *IEEE 14-Bus System*: A commonly used benchmark in power system analysis, featuring 14 buses, 5 generators, and 11 loads connected via 20 lines [20]. It offers a moderately complex environment suitable for assessing DRL agents in discrete AVC tasks.

2. *Illinois 200-Bus System*: A synthetic model with higher complexity, consisting of 200 buses distributed over six areas, equipped with 49 generators, and 160 loads [21]. It challenges DRL agents with its 179 lines and 66 transformers, testing their performance in continuous AVC scenarios.

3. *ISO-NE Node-Breaker Model*: The most complex system in our study includes a node-breaker configuration of ~21,000 buses, ~24,000 branches, and ~19,000 breakers. Simplifying the system by merging buses connected exclusively by breakers, the analysis is streamlined to focus on ~1,200 buses and ~3,400 branches, representing the core elements for evaluation.

Latin Hypercube Sampling (LHS) was used to modify base loads and generation profiles, ensuring a comprehensive range of operational conditions for our tests, and enhancing the robustness of our assessment [22].

*B. Model Fidelity and Agent Performance*

One concern from the system operators is whether the trained agent can handle model inaccuracy, which is inevitable for real systems. To address this, we investigate the influence of model fidelity on the performance of DRL agents in operational power systems. Our approach deliberately introduces variable impedance parameter inaccuracies during the training phase within the IEEE 14-bus and Illinois 200-bus system simulations. However, for testing, we employ accurate grid models to assess the "real-world" applicability of the trained agents.

We adopt a robust fivefold cross-validation framework, incorporating unique random seeds in each fold to foster diverse learning conditions while preserving a consistent test environment. The validation involves 15,000 simulation episodes, including 10,000 training and 5,000 testing episodes, to thoroughly evaluate the voltage control adeptness of DRL agents under various fidelity scenarios.

The core of the study evaluates the consequences of deploying DRL agents, which were trained on flawed grid models, in real-time power grid operations. By simulating discrepancies and contrasting them against accurate models in testing, we aim to quantify their impact on agent efficacy and reinforce the reliability of AI in autonomous grid management. Our detailed examination highlights the DRL algorithms' challenges in handling complex network conditions. We maintain uniformity in network architecture and hyperparameters across four distinct testing conditions for each algorithm to enable fair comparison: 1) a precise grid model, 2) an 8% impedance error on a single line, 3) a 20% impedance error on a single line, and 4) random impedance errors ranging from -20% to +20% on all lines.

Fig. 2 conveys the Illinois 200-bus system's performance metrics, with average rewards per 50 episodes illustrating the trade-off between control iteration count and reward efficiency. A reward of 400, denoting minimal control effort, marks the benchmark for optimal policy execution. Conversely, increasing iteration counts for voltage regulation correlates with a reduced reward score, indicating performance degradation. Despite these challenges, the DDPG agents exhibit remarkable resilience by achieving near-optimal policy outcomes during the testing phases, signifying their proficiency even when preliminary training involves imprecise model data.

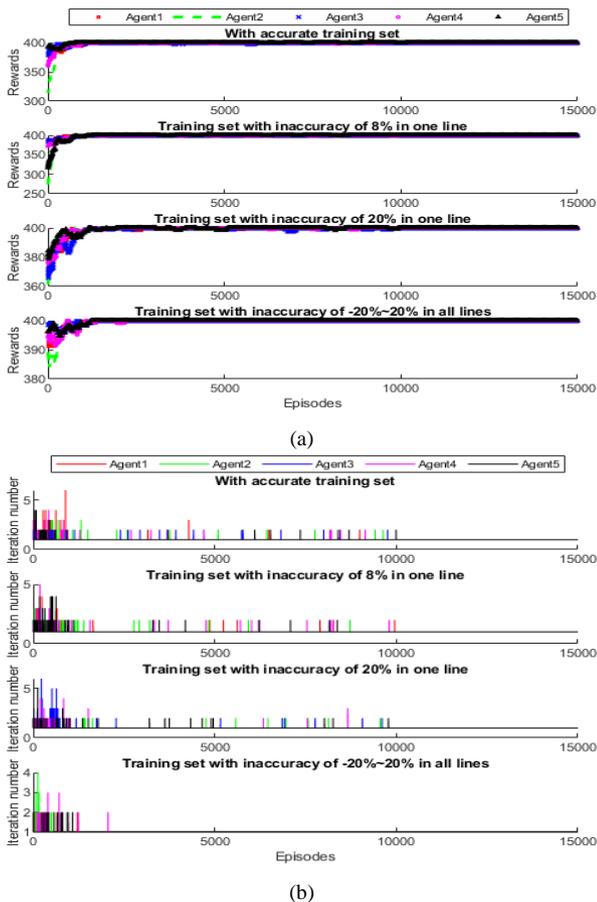

Fig. 2 (a) The Illinois 200-bus rewards achieved by the DDPG agent without loss consideration, (b) corresponding control iterations needed.

More experiments are conducted where we augment the agent's reward function (1) with additional terms, $F_p$ and $F_n$, to include system losses, aiming to optimize overall system efficiency beyond voltage regulation:

$$F_p = F_n = -\varepsilon \cdot L, \quad (9)$$

where $\varepsilon$ is a scaling factor and $L$ denotes the system total power loss. Reflecting the findings of Fig. 2, this enhancement similarly affected the agent's performance. A comparative analysis of system losses during the testing phase highlights these effects. Fig. 3 elucidates our findings by charting the incremental losses—quantified as the deviation from a scenario utilizing an exact grid model—across successive episodes. The inclusion of this figure underscores the effectiveness of the adapted reward function in enhancing system efficiency, even when the model undergoes variations. The modification of the reward function has proven effective in maintaining operational efficacy across various grid conditions.

Our study confirms that DRL agents can effectively perform voltage control tasks, even in the presence of grid model inaccuracies, thereby proving their suitability for real-world power grid management. The agents reliably maintain safe voltage levels and operational efficiency, demonstrating the potential for AI-driven system operations. Results show that DRL agents, such as those using the DDPG framework, quickly correct voltage violations and achieve optimal performance within a single iteration, as illustrated in Fig. 2.

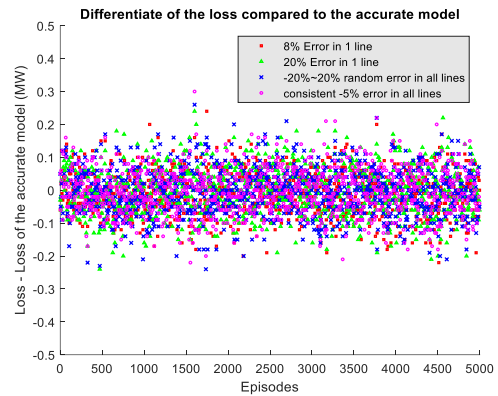

Fig. 3. Comparison of system loss in the testing phase of each scenario.

Although model inaccuracies may impact loss minimization, they do not significantly compromise the agents' overall effectiveness. The consistent performance across the 14-bus and 200-bus systems substantiates the robustness of DRL strategies, providing valuable evidence for system operators considering the adoption of AI technologies.

### C. Discrete vs. Continuous Control Paradigms

In this subsection, we examine the performance of discrete and continuous control strategies for voltage control in a heavily loaded area of the Illinois system. We compare the abilities of two DRL agents, DQN and DDPG, across 10,000 training episodes and 4,000 test scenarios to evaluate their efficiency and adaptability under varying conditions. The DQN operates with a fixed action set, offering 625 distinct voltage set points for generators after discretizing the action space. In contrast, the DDPG benefits from a continuous action space, allowing for granular control actions.

Fig. 4 highlights the specific load center under study, while Fig. 5 and Fig. 6 compare the performance of DQN and DDPG, respectively. Our analysis confirms that DDPG surpasses DQN in terms of convergence speed and adaptability, which is particularly evident during *N*-1 contingency training—a finding consistent with previous observations on the IEEE 14-bus system.

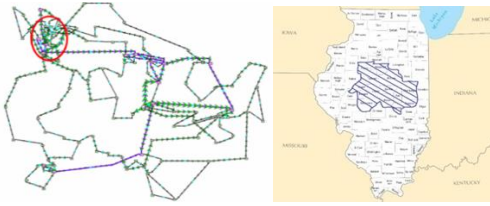

Fig. 4. A load center of the 200-bus model selected for testing DRL agents.

Despite their differences, both agents demonstrate a notable degree of robustness and capacity for autonomous learning. DQN is better suited to simpler, lower-dimensional environments, whereas DDPG is more apt for complex scenarios that require a detailed, continuous range of actions, at the cost of higher computational load and sensitivity to parameter tuning.

For both agents, most of their training time is spent exchanging data with the power flow solver, resulting in similar training durations of roughly 18 minutes for the 14-bus system and 25 minutes for the 200-bus system on a standard computing setup.

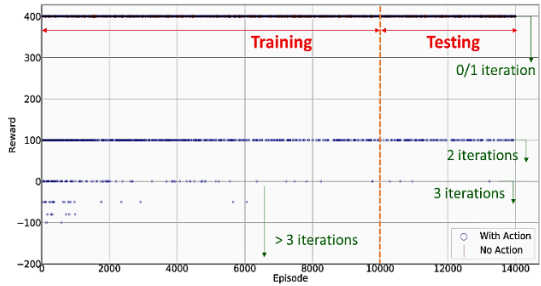

Fig. 5. DQN agent performance on the Illinois 200-bus system.

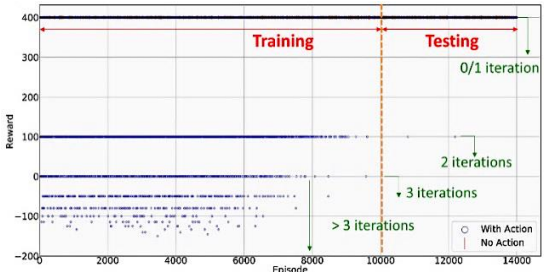

Fig. 6. DDPG agent performance on the Illinois 200-bus system.

In anticipation of future work, we identify the SAC algorithm as a promising successor to DDPG for DRL in power systems. Backed by numerous case studies, SAC's entropy-based policy promotes exploratory action, enhancing sample efficiency and yielding a more robust training outcome, as compared to DDPG. Accordingly, we will focus on SAC in forthcoming case studies to maintain conciseness.

### D. Topological Variability and DRL Agent Adaptability

Another system operators' concern is whether the agents' actions are still valid after topological changes in power networks, e.g., $N$-1 line outages. This subsection explores the resilience and adaptability of DRL agents within the Illinois and the ISO-NE model—two test cases designed to capture the complexity of real-world grids.

The Illinois 200-bus scenario examines the adaptability of DRL agents when facing disruptions caused by disconnecting the ten most heavily loaded lines: 187-121, 14-121, 188-89, 194-150, 83-146, 55-102, 102-128, 14-149, 123-133, and 81-55. Fig. 7 provides an in-depth analysis of the SAC agent' robustness during these targeted outages, showcasing performance metrics through 20,000 training instances and 10,000 testing scenarios.

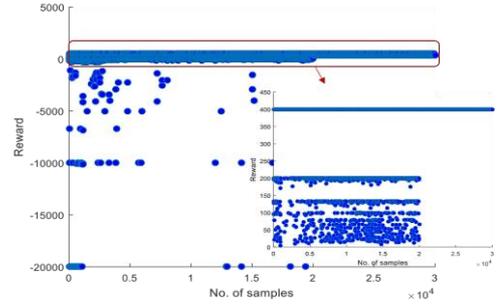

Fig. 7. SAC agent performance on 200-bus system with random $N$-1.

In tandem, the ISO-NE model was used to evaluate the agents' adaptability to 20 selected $N$-1 and even $N$-2 contingencies. The results of this evaluation are depicted in Fig. 8. It is noteworthy that a reward score of 500 signifies the successful mitigation of voltage violations through line disconnections, prior to the agent's intervention.

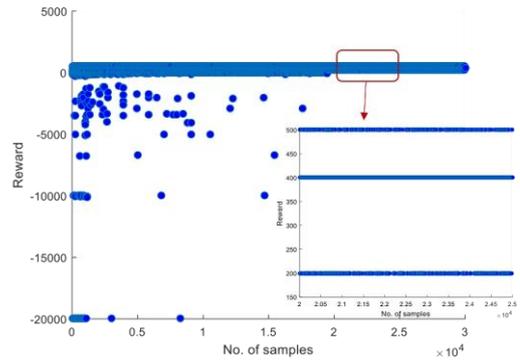

Fig. 8. SAC agent performance on ISO-NE system with random $N$-1.

The collective analysis from these studies highlights the agents' resilience and their capacity to maintain grid security after major contingencies. Accordingly, these outcomes suggest that incorporating DRL agents into the operational framework of power systems can enhance the grid's robustness against unpredictable events.

### E. Agent's Performance under Unseen Scenarios

The last but not the least of system operators' concern is whether the trained agents can still perform on a slightly changed system. This aspect is critical to the reliability of these agents when they encounter real-world contingencies beyond their programmed experience. In response to this concern, we present an empirical analysis of a DRL agent's performance against unencountered operational disruptions. Using the SAC agent trained for the ISO-NE system in *subsection D*, we assess its generalization capacity with a set of 10 new $N$-1/$N$-2 contingencies over 5,000 test scenarios.

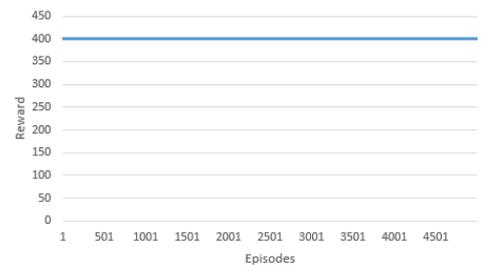

Fig. 9. SAC agent performance on 10 new $N$-1/$N$-2 contingencies.

The contingencies were selected for their variety and the degree of challenge they pose, aiming to test the agent's response to unfamiliar and potentially stressful operational changes. Fig. 9 displays the agent's performance in managing these contingencies. The results are noteworthy: the agent consistently and successfully mitigated voltage violations in one control iteration across all new test cases. This efficiency in handling unforeseen events reinforces the DRL agent's value in improving the resilience of power systems, proving its ability to adapt to significant topological changes without prior direct training.

*F. Minimum Control Efforts*

Besides the generalization requirements, system operators also prioritize efficient interventions for voltage control, specifically aiming to conserve resources during over-voltage or under-voltage events—a challenge not yet fully explored in existing studies. To align with these practical concerns, our research modifies the reward function in the DRL agent's algorithm to incentivize minimal control action. We introduce an efficiency term, $\varepsilon$, to equation (10)

$$F_p = F_n = \varepsilon \cdot N; \quad (10)$$

where $N$ reflects the count of unchanged equipment, reflecting a strategic approach to resource conservation. This innovation underscores a targeted application for load tap changing (LTC) transformers within a 200-bus power system, promoting an operational philosophy where fewer interventions are preferred, thereby aligning closer with the practical needs of system operation.

TABLE 1 RESULTS FOR TESTING CASES WITH DIFFERENT $\varepsilon$ – NUMBER OF CASES WITH SPECIFIED NUMBERS OF LTC ACTIONS

| Agent no. | 17-N 17 | 16 | 15 | 14 | 13 | 12 | 11 | 10 | 9 | 8 | 7 | 6 | 5 |
|---|---|---|---|---|---|---|---|---|---|---|---|---|---|
| Agent 1 ($\varepsilon = 0$) | 2605 | 7223 | 124 | 48 | 0 | 0 | 0 | 0 | 0 | 0 | 0 | 0 | 0 |
| Agent 2 ($\varepsilon = 0$) | 2174 | 6224 | 1419 | 179 | 4 | 0 | 0 | 0 | 0 | 0 | 0 | 0 | 0 |
| Agent 3 ($\varepsilon = 0$) | 490 | 3161 | 4967 | 1132 | 65 | 27 | 158 | 0 | 0 | 0 | 0 | 0 | 0 |
| Agent 1 ($\varepsilon = 2$) | 0 | 80 | 137 | 579 | 1150 | 1279 | 534 | 3568 | 1515 | 1044 | 114 | 0 | 0 |
| Agent 2 ($\varepsilon = 2$) | 0 | 134 | 1967 | 1539 | 817 | 484 | 1397 | 2649 | 1013 | 0 | 0 | 0 | 0 |
| Agent 3 ($\varepsilon = 2$) | 0 | 464 | 4735 | 3192 | 1350 | 259 | 0 | 0 | 0 | 0 | 0 | 0 | 0 |
| Agent 1 ($\varepsilon = 3$) | 0 | 0 | 0 | 0 | 0 | 0 | 148 | 322 | 329 | 611 | 879 | 1858 | 1986 |
| Agent 2 ($\varepsilon = 3$) | 0 | 0 | 0 | 0 | 503 | 967 | 7259 | 901 | 370 | 0 | 0 | 0 | 0 |
| Agent 3 ($\varepsilon = 3$) | 0 | 0 | 0 | 240 | 434 | 1147 | 1841 | 3339 | 1724 | 1157 | 74 | 44 | 0 |

The training of the agent on 20,000 scenarios was specifically designed to prioritize efficiency in decision-making, aligning with the priorities of system operators. Testing the agent against 10,000 different scenarios with varying random seeds, we observed strict adherence to minimal control actions as dictated by the efficiency variable $\varepsilon$. The findings presented in Table I affirm the agent's utility in aligning with pragmatic system operation objectives. For instance, with $\varepsilon$ set to 0, the data indicates that 2,605 out of 10,000 cases required movement of all 17 LTCs for voltage control. As $\varepsilon$ increased, the number of LTC adjustments decreased, showcasing that reward shaping can effectively reduce the necessity for control actions while ensuring reliable voltage control.

## IV. CONCLUSION AND FUTURE WORK

This study marks a significant advancement in aligning DRL-based AVC with the practical needs of system operators, by rigorously testing its efficacy in realistic grid conditions using various system models. Our findings underscore the adaptability of DRL in actual grid management and offer valuable insights into its practical deployment challenges.

Future work will focus on bridging the gap between simulation and operation, enhancing DRL algorithms for better real-world application. We will explore DRL's robustness in extreme grid situations such as extreme penetrations of renewables. Ongoing efforts will also address optimizing dispatch and control for safety, cost, and efficiency. Success in these endeavors calls for multidisciplinary collaboration, integrating insights from regulation, economy, and engineering to develop AI-driven solutions that are both effective and sustainable.

## V. REFERENCES

[1] J. Duan, D. Shi, et al., "Deep-Reinforcement-Learning-Based Autonomous Voltage Control for Power Grid Operations," *IEEE Trans. Power Syst.*, vol. 35, no. 1, pp. 814-817, Jan. 2020.
[2] M. M. Hosseini and M. Parvania, "Resilient Operation of Distribution Grids Using Deep Reinforcement Learning," *IEEE Trans. Industrial Informatics*, vol. 18, no. 3, pp. 2100-2109, Mar. 2022.
[3] R. Huang, Y. Chen, et al., "Learning and Fast Adaptation for Grid Emergency Control via Deep Meta Reinforcement Learning," *IEEE Trans. Power Syst.*, vol. 37, no. 6, pp. 4168-4178, Nov. 2022.
[4] Y. Ye, D. Papadaskalopoulos, et al., "Multi-Agent Deep Reinforcement Learning for Coordinated Energy Trading and Flexibility Services Provision in Local Electricity Markets," *IEEE Trans. Smart Grid*, vol. 14, no. 2, pp. 1541-1554, Mar. 2023.
[5] L. Yan, X. Chen, et al., "A Hierarchical Deep Reinforcement Learning-Based Community Energy Trading Scheme for a Neighborhood of Smart Households," *IEEE Trans. Smart Grid*, vol. 13, no. 6, pp. 4747-4758, Nov. 2022.
[6] T. Chen, S. Bu, et al., "Peer-to-Peer Energy Trading and Energy Conversion in Interconnected Multi-Energy Microgrids Using Multi-Agent Deep Reinforcement Learning," *IEEE Trans. Smart Grid*, vol. 13, no. 1, pp. 715-727, Jan. 2022.
[7] H. Xu, Z. Yu, et al., "Improved deep reinforcement learning based convergence adjustment method for power flow calculation*," The 16th IET International Conference on AC and DC Power Transmission (ACDC 2020)*, pp. 1898-1903, 2020.
[8] Y. Liang, Z. Ding, et al., "Real-Time Operation Management for Battery Swapping-Charging System via Multi-Agent Deep Reinforcement Learning," *IEEE Trans. Smart Grid*, vol. 14, no. 1, pp. 559-571, Jan. 2023.
[9] J. Jin, S. Mao and Y. Xu, "Optimal Priority Rule Enhanced Deep Reinforcement Learning for Charging Scheduling in an Electric Vehicle Battery Swapping Station," *IEEE Trans. Smart Grid*, early access, Feb. 2023.
[10] A. A. Amer, K. Shaban, and A. M. Massoud, "DRL-HEMS: Deep Reinforcement Learning Agent for Demand Response in Home Energy Management Systems Considering Customers and Operators Perspectives," *IEEE Trans. Smart Grid*, vol. 14, no. 1, pp. 239-250, Jan. 2023.
[11] S. Nematshahi, D. Shi, et al., "Deep Reinforcement Learning Based Voltage Control Revisited," *IET Gener., Trans. and Distr.*, vol. 17, no. 21, pp. 4826-4835, Oct. 2023.
[12] L. Ding, Z. Lin, et al., "Target-Value-Competition-Based Multi-Agent Deep Reinforcement Learning Algorithm for Distributed Nonconvex Economic Dispatch," *IEEE Trans. Power Syst.*, vol. 38, no. 1, pp. 204-217, Jan. 2023.
[13] X. Liu and C. Konstantinou, "Reinforcement learning for cyber-physical security assessment of power systems," *Proc. of the 2019 IEEE Milan PowerTech Conf.*, 2019.
[14] Y. Wang and B. Pal, "Destabilizing Attack and Robust Defense for Inverter-Based Microgrids by Adversarial Deep Reinforcement Learning," *IEEE Trans. Smart Grid*, early access, March 2023.
[15] T. T. Nguyen and V. J. Reddi, "Deep Reinforcement Learning for Cyber Security," *IEEE Trans. Neural Networks and Learning Systems*, early access, Nov 2022.
[16] T. Lan, J. Duan, et al., "AI-based autonomous line flow control via topology adjustment for maximizing time-series ATCs," *IEEE PES General Meeting*, pp. 1-5, 2020.
[17] D. Silver, T. Hubert, et al., "Mastering chess and shogi by self-play with a general reinforcement learning algorithm," *arXiv preprint arXiv:1712.01815*, 2017.
[18] P. Xu, J. Duan, et al., "Active Power Correction Strategies Based on Deep Reinforcement Learning—Part I: A Simulation-driven Solution for Robustness," *CSEE Journal of Power and Energy Systems*, vol. 8, no. 4, pp. 1122-1133, July 2022.
[19] S. Wang, J. Duan, et al., "A Data-Driven Multi-Agent Autonomous Voltage Control Framework Using Deep Reinforcement Learning," *IEEE Trans. Power Syst.*, vol. 35, no. 6, pp. 4644-4654, Nov. 2020.
[20] IEEE 14-bus System. [Online]. Available: https://icseg.iti.illinois.edu/ieee-14-bus-system/.
[21] Illinois 200-bus System: ACTIVSg200. [Online]. Available: https://electricgrids.engr.tamu.edu/electric-grid-test-cases/activsg200/.
[22] K. Q. Ye, "Orthogonal Column Latin Hypercubes and their Application in Computer Experiments," *J. Am. Stat. Assoc.*, vol. 93, no. 444, pp. 1430-1439, 1998.